\documentclass[a4paper]{article}
\usepackage{Odyssey2018}
\usepackage{amssymb, amsmath, bm, mathtools,array}
\usepackage{hyperref}
\usepackage{url}
\usepackage{multirow}
\usepackage{verbatim}
\usepackage{textcomp}
\usepackage{subfig}
\ninept

\newcommand{\bs}[1]{\boldsymbol{#1}}
\DeclareMathOperator{\EX}{\mathbb{E}}

\title{Can we steal your vocal identity from the Internet?: Initial investigation of cloning Obama's voice using GAN, WaveNet and low-quality found data}

\makeatletter
\def\name#1{\gdef\@name{#1\\}}
\makeatother
\name{{\em Jaime Lorenzo-Trueba$^1$, Fuming Fang$^1$, Xin Wang$^1$, Isao Echizen$^1$}, \\{\em Junichi Yamagishi$^{1,2}$, Tomi Kinnunen$^3$}}
\address{
$^1$ National Institute of Informatics, Tokyo, Japan 
$^2$ University of Edinburgh, Edinburgh, UK \\
$^3$ University of Eastern Finland, Joensuu, Finland \\
{\small \tt \{jaime, fang, wangxin, iechizen, jyamagis\}@nii.ac.jp, tkinnu@cs.uef.fi}}
\begin{document}
\maketitle
\begin{abstract}

Thanks to the growing availability of spoofing databases and rapid advances in using them, systems for detecting voice spoofing attacks are becoming more and more capable, and error rates close to zero are being reached for the ASVspoof2015 database. However, speech synthesis and voice conversion paradigms that are not considered in the ASVspoof2015 database are appearing. Such examples include direct waveform modelling and generative adversarial networks. We also need to investigate the feasibility of training spoofing systems using only low-quality found data. For that purpose, we developed a generative adversarial network-based speech enhancement system that improves the quality of speech data found in publicly available sources. Using the enhanced data, we trained state-of-the-art text-to-speech and voice conversion models and evaluated them in terms of perceptual speech quality and speaker similarity. The results show that the enhancement models significantly improved the SNR of low-quality degraded data found in publicly available sources and that they significantly improved the perceptual cleanliness of the source speech without significantly degrading the naturalness of the voice. 
However, the results also show limitations when generating speech with the low-quality found data.

\end{abstract}

\section{Introduction}
\label{sec:intro}

ASVspoof2015~\cite{7858696} is a commonly used database for developing and evaluating methods for preventing spoofing of automatic speaker verification (ASV) systems. Recent studies have shown very good spoofing detection rates for the ASVspoof2015 corpus with equal error rates (EERs) close to zero~\cite{muckenhirn2017end}. 

However, with recent advances in text-to-speech (TTS) and voice conversion (VC) techniques in the speech synthesis field, it has become clear that the TTS and VC systems included in the corpus are out of date. Good examples of state-of-the-art speech synthesis techniques include direct waveform modeling~\cite{oord2016wavenet,Wang2017,Tamamori2017} and generative adversarial networks (GANs)~\cite{kaneko2017generative}. In a recent study~\cite{46619}, speech synthesized using an end-to-end direct waveform model was rated as natural as human speech. A preliminary study showed that direct waveform models do not fool speaker recognition systems; however, Mel-spectra generated by a GAN have successfully fooled a speaker recognition system~\cite{wilson2018attacking}\footnote{But the experiment was not fair as the attackers had knowledge of the features used in the recognition system and directly input generated features to the system, which are impractical assumptions.}.

One drawback of the studies using the ASVspoof2015 corpus is that they used an unrealistic assumption: only studio-recorded speech was used for generating spoofing materials. If we think of more potential spoofing scenarios, it is reasonable to expect attacks using unwillingly obtained speech (i.e., recordings made in non-controlled environments). This motivated us to study the feasibility of training a speech enhancement system by obtaining publicly available data and then using tools to enhance the speech quality of the found data with the aim of creating reasonable state-of-the-art spoofing TTS or VC materials.

In the case of using publicly available found data for generating spoofed speech, it is easy to understand that there are large amounts of speech data in publicly available sources and that it is likely that data for almost anybody can be found one way or another. If we are talking about public personalities like President Barack Obama (a common target for identity theft research~\cite{kumar2017obamanet,suwajanakorn2017synthesizing}), the amount of data publicly available is immense.

Such data is commonly recorded in non-professional acoustic environments such as homes and offices. Moreover, the recordings are often made using consumer devices such as smartphones, tablets, and laptops. Therefore, the speech portions of the recordings are typically of poor quality and contain a large amount of ambient noise and room reverberation. However, applications developed for speaker adaptation of speech synthesis and for voice conversion have been normally designed to work only on clean data with optimal acoustic quality and properties. Therefore, the quality of systems trained using data found in publicly available sources is unknown.

Our specific objective was to answer two questions. First, how well can we train a speech enhancement system to enhance low-quality data found in publicly available sources? Second, can we use such enhanced data to produce effective spoofing materials using the best available TTS and VC systems?

\section{GAN-based speech enhancement}
\label{sec:senhc}

Generative adversarial networks consist of two ``adversarial'' models: a generative model $G$ that captures the data distribution and a discriminative model $D$ that estimates the probability that a sample came from the training data rather than $G$. This GAN structure has been used to enhance speech~\cite{pascual2017segan,donahue2017exploring}. In the research reported here, we worked on improving the speech enhancement generative adversarial network (SEGAN)~\cite{pascual2017segan}. More specifically, we attempted to make the training process more robust and stable by introducing a modified training strategy for SEGAN's generator.

\subsection{SEGAN-based speech enhancement}\label{ssec:rsegan}

SEGAN~\cite{pascual2017segan} exploits the generative adversarial structure in a particular fashion. The speech enhancement itself is carried out mainly by using model $G$, which follows an encoder-decoder structure that takes noisy speech as an input and produces enhanced speech as the output, similar to the U-net architecture used in the pix2pix framework~\cite{isola2017image}. The function of model $D$ is to determine, during training, whether the enhanced speech is detected as fake (enhanced) or real (clean). 
If model $D$ can be fooled by the enhanced speech, there is no gradients through model $G$. On the other hand, if model $D$ cannot be fooled by the enhanced speech, the gradient is back-propagated through the model $G$ and update it in order to fool the model $D$, thus biasing the structure towards producing enhanced speech closer to the clean speech.

We found that the SEGAN structure is sensitive to noise variations during training, making convergence difficult. We thus made two modifications to achieve more robust training. For the first one, we created some pre-trained baseline speech enhancement models (which may be simpler signal processing methods or easier-to-train neural networks than GANs). They are used to enhance the speech, and then the content loss of the initial iterations of the generator model is computed on the basis of the baseline enhanced speech instead of on clean speech.

For the second modification, a skip connection was added around the generator so that its task is not to generate enhanced speech from scratch but to generate a residual signal that refines the input noisy speech~\cite{kaneko2017generative}. This should encourage the generator to learn the detailed differences between clean and enhanced speech waveforms.

\section{Speech Corpora}
\label{sec:corpus}

We used two types of speech corpora. The corpora used to train the speech enhancement module were constructed using publicly available data so that the training process would be replicable. The corpus used as the source for the cloned voice was constructed using a number of President Barack Obama's public interventions obtained from various sources.

\subsection{Corpora for speech enhancement training}\label{ssec:secorpus}

For training the speech enhancement module, we used a subset (28 speakers; 14 male and 14 female; British accent; ~400 utterances per speaker) of the Centre for Speech Technology Research (CSTR) voice cloning toolkit (VCTK) corpus\footnote{http://datashare.is.ed.ac.uk/handle/10283/1942}~\cite{veaux2013voice} as the clean speech corpus. We used different noisy iterations of this corpus to create four additional corpora for use in making the speech enhancement signal robust against noisy and/or reverberant environments (see table~\ref{tab:SEsources}). These corrupted corpora were recorded as a collaboration between CSTR and the National Institute of Informatics of Japan and are publicly available in the DataShare repository of the University of Edinburgh.

\begin{table}[!t]
\centering
\caption{Corpora for speech enhancement training.}
\vspace{-3mm}
\begin{tabular}{|l|l|l|l|}\hline
Corpus Name & Abbrev. & \#Files & Total Time \\ \hline
VCTK & clean & 11572 &  8h54m56s\\
Noisy VCTK & n & 11572 & 8h54m56s \\
Reverberant VCTK & r & 11572 & 8h54m56s \\
Noisy Reverberant VCTK & nr & 11572 & 8h54m56s \\
Device Recorded VCTK & DR & 11572 & 8h54m56s \\\hline
\end{tabular}
\label{tab:secorpus}
\end{table}

\subsubsection{Device-recorded VCTK corpus}

To create the device-recorded (DR) VCTK corpus\footnote{https://datashare.is.ed.ac.uk/handle/10283/2959}~\cite{sarfjoo2017device} we re-recorded the high-quality speech signals in the original VCTK corpus by playing them back and recording them in office environments using relatively inexpensive consumer devices. This corpus enables our speech enhancement system to learn the nuanced relationships between high quality and device-recorded versions of the same audio. Eight different microphones were used simultaneously for the re-recording, which was carried out in a medium-sized office under two background noise conditions (i.e., windows either opened or closed). This resulted in 16 different conditions.

\subsubsection{Noisy, Reverberant, and Noisy and reverberant VCTK corpora}

We used three other artificially corrupted variations of the CSTR VCTK corpus: Noisy VCTK\footnote{https://datashare.is.ed.ac.uk/handle/10283/2791}~\cite{Valentini-Botinhao+2016,Valentini-Botinhao+SSW2016}, Reverberant VCTK\footnote{https://datashare.is.ed.ac.uk/handle/10283/2031}~\cite{valentini2016reverberant}, and Noisy and reverberant VCTK\footnote{https://datashare.is.ed.ac.uk/handle/10283/2826}~\cite{valentini2017noisyrev}. This diverse portfolio of possible speech corruptions enables our speech enhancement system to learn how to target different possibilities, i.e., plain noisy, reverberation compensation, and a mixture of both. As mentioned above, all the corpora are based on the CSTR VCTK corpus, so the speakers and utterances represented in the Edinburgh noisy speech dataset are similar to those of the DR-VCTK corpora.

\subsection{Obama's found data}\label{ssec:obamacorpus}

Obama's data was found online, mainly in YouTube videos with transcriptions as part of the description and from diverse sources such as interviews and political meetings. The recording conditions and environments were diverse, ranging from very noisy with large amounts of reverberation to not so noisy or not so reverberant, and never achieving recording studio standards. The audio channel was split from the video channel, automatically segmented on long pauses, and down-sampled to 16 kHz. The transcription was copied over as a text file. Table~\ref{tab:obamadesc} shows a brief characterization of the data.

\begin{table}[!t]
\centering
\caption{Characterization of the used Obama's found data.}
\vspace{-3mm}
\begin{tabular}{|l|l|}\hline
Sources & Public speeches, interviews... \\
Total length (w. silence) & 3h 7m 39s \\
Minimum segment duration & 0.54s \\
Maximum segment duration & 24.4s \\
Average segment duration & 5.4s \\
Estimated SNR mean & 17.15 dB \\
Estimated SNR variance & 171.22 dB \\ \hline
\end{tabular}
\label{tab:obamadesc}
\vspace{-3mm}
\end{table}

\begin{figure}[tb]
	\centering
	\includegraphics[width=0.45\textwidth]{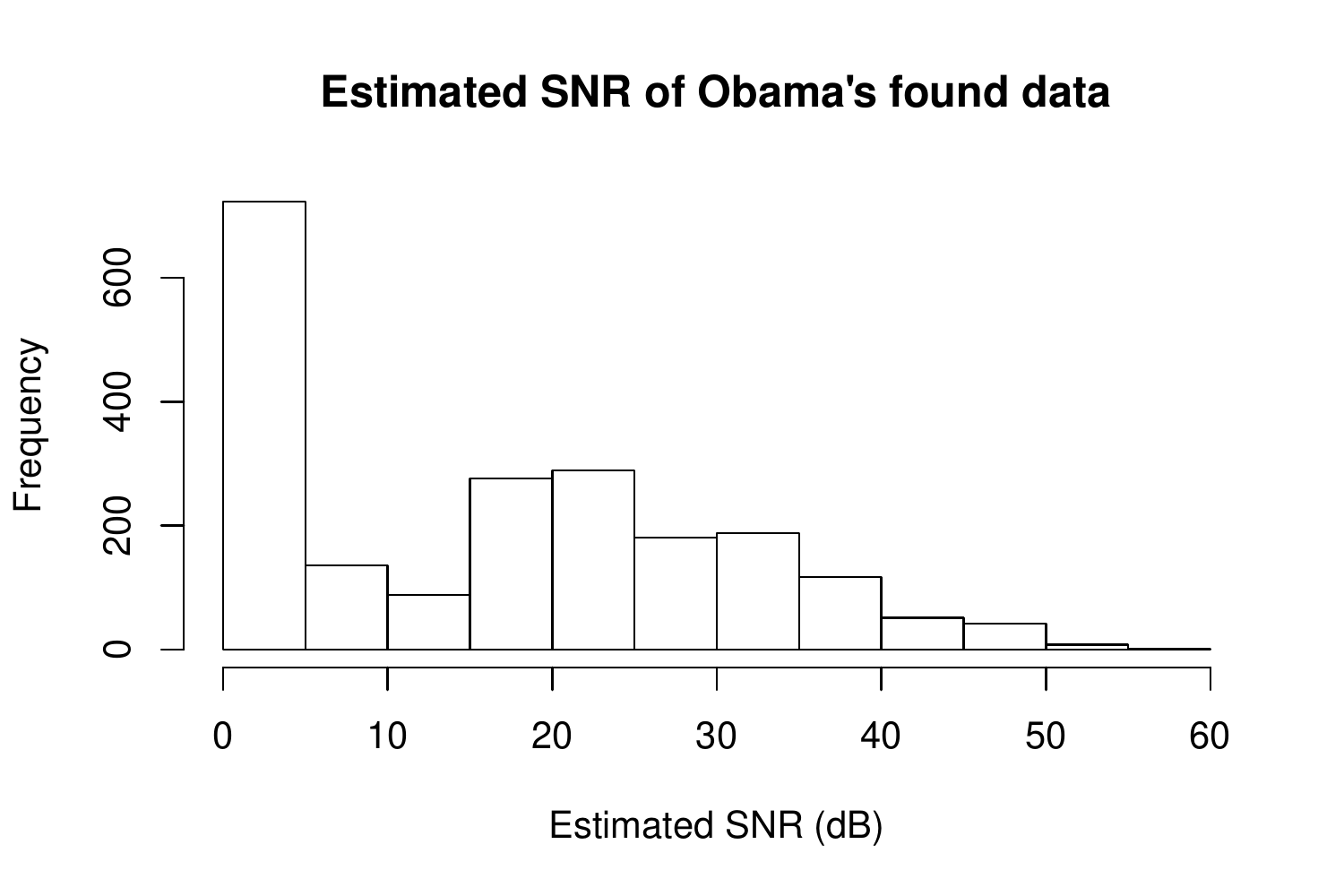}
	\vspace{-5mm}
	\caption{SNR histogram of original Obama's found data.}
	\label{fig:snrhist}
\end{figure}

A histogram of the signal-to-noise ratio (SNR) in dB, estimated using the NIST SNR measurement tool \footnote{https://www.nist.gov/information-technology-laboratory/iad/mig/nist-speech-signal-noise-ratio-measurements} is shown in figure~\ref{fig:snrhist}. It is evident that the vast majority of the speech signals had a very low SNR compared with conventional speech generation corpus standards. For instance, the mean and variance of the SNRs estimated for the VCTK corpus were 30.19 and 23.73 dB, respectively, whereas those of the data used as the source for the cloned voice were 17.15 and 171.22 dB.

\section{Enhancing Obama's found data}\label{sec:enhcobamac}

As mentioned above, the aim of this research was to create reasonable state-of-the-art spoofing TTS or VC materials. More specifically, we aimed to train a waveform generation model that can replicate the target speaker’s voice, in this case, the very recognizable voice of President Barack Obama. Moreover, the training should be done using only easily available low-quality resources, as explained in section~\ref{ssec:obamacorpus}. But that kind of data is generally too poor to ensure reasonably good training of speech synthesis systems. We thus developed a generative adversarial network-based speech enhancement system for improving low-quality data to the point where it can be used to properly train TTS systems.

\subsection{Design of the speech enhancement models}

As we had a large amount of free, publicly available resources for training our speech enhancement system, we did testing to determine the best training regime strategy. We trained our speech enhancement system using seven different sources with various amounts of data, as summarized in table~\ref{tab:SEsources}.

\begin{table}[!t]
\centering
\caption{Data sources used for training speech enhancement model. (See table~\ref{tab:secorpus} for the meanings of the abbreviations.)}
\vspace{-3mm}
\begin{tabular}{|l|l|l|}\hline
SOURCES & \#Files & Total Time \\ \hline
DR & 11572 & $~~$8h54m56s \\
n & 11572 & $~~$8h54m56s \\
r & 11572 & $~~$8h54m56s \\
nr & 11572 & $~~$8h54m56s \\
DR+n & 23144 & 17h49m52s \\
DR+nr & 23144 & 17h49m52s \\
All (DR+n+r+nr) & 46288 & 35h39m44s\\\hline
\end{tabular}
\label{tab:SEsources}
\vspace{-3mm}
\end{table}

Our motivation for using three single-category sources and four combinations was our expectation that training using each single-category source would work well for the corresponding type of disturbance (i.e., training using source ``n'' should be good for cleaning noise, ``r'' should be good for cleaning reverberation, and ``DR'' should be good for compensating for low-quality recording devices). Since most of the data found will come from noisy poor-quality sources, it made sense to combine ``DR'' with the different noisy corpora. Moreover, since having as much varied data as possible helps neural networks generalize better, the combination of all the corpora should also be effective.

\subsection{Training of the speech enhancement models}

Similar to the original SEGAN training strategy, we extracted chunks of waveforms by using a sliding window of $2^{14}$ samples at every $2^{13}$ samples (i.e., 50\% overlap). At testing time, we concatenated the results at the end of the stream without overlapping. For the last chunk, instead of zero padding, we pre-padded it with the previous samples. For batch optimization, RMSprop with a 0.0002 learning rate and a batch size of 100 was used. The modified SEGAN model converged at 120 epochs.

For selecting the pre-enhancement method, we conducted preliminary experiments, using the Postfish~\cite{postfish2005} and the HRNR~\cite{plapous2006improved}. Their sequential use enhanced the quality of the samples, so we used this compound method to generate the baseline models described in section~\ref{ssec:rsegan}.

The source code and extended documentation for the SEGAN implementation are available online\footnote{https://github.com/ssarfjoo/improvedsegan}.

\subsection{Objective evaluation of the speech enhancement}\label{ssec:objective}

After the speech enhancement models were trained, they were applied to the noisy data. The effect of the enhancement process was evaluated by estimating the SNR once again using the NIST tool. Although SNR is most likely not the best measure of enhancement, the lack of a clean reference limited the availability of tools.

\begin{table}[t!]
\centering
\caption{Average SNR in dB estimated with NIST tool for the results of the different speech enhancement models.}
\vspace{-3mm}
\begin{tabular}{|l|l|}\hline
SOURCES & average SNR (dB) \\ \hline
Obama source & 17.2 \\
n & 49.8 \\
r & 22.7 \\
nr & 43.1 \\
DR & 28.24 \\
DR+n & 40.1 \\
DR+nr & 41.37 \\
all (DR+n+r+nr) & 37.89 \\ \hline
\end{tabular}
\label{tab:ObjResults}
\vspace{-5mm}
\end{table}

\begin{figure}[tb]
	\centering
	\includegraphics[width=0.45\textwidth]{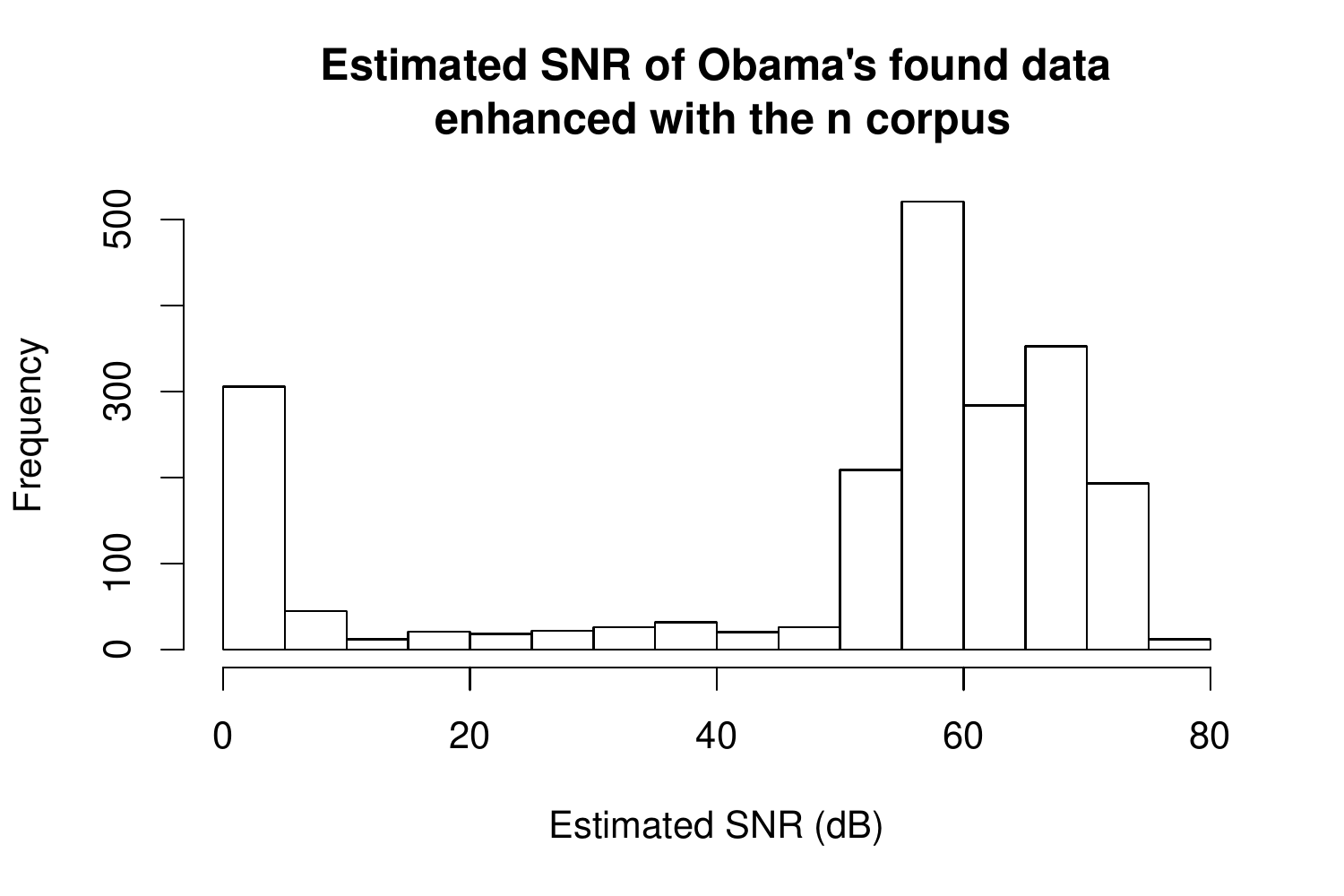}
	\vspace{-5mm}
	\caption{SNR histogram of original Obama voice data after enhancement using noisy VCTK.}
	\label{fig:nhist}
\end{figure}

\begin{figure}[tb]
	\centering
	\includegraphics[width=0.45\textwidth]{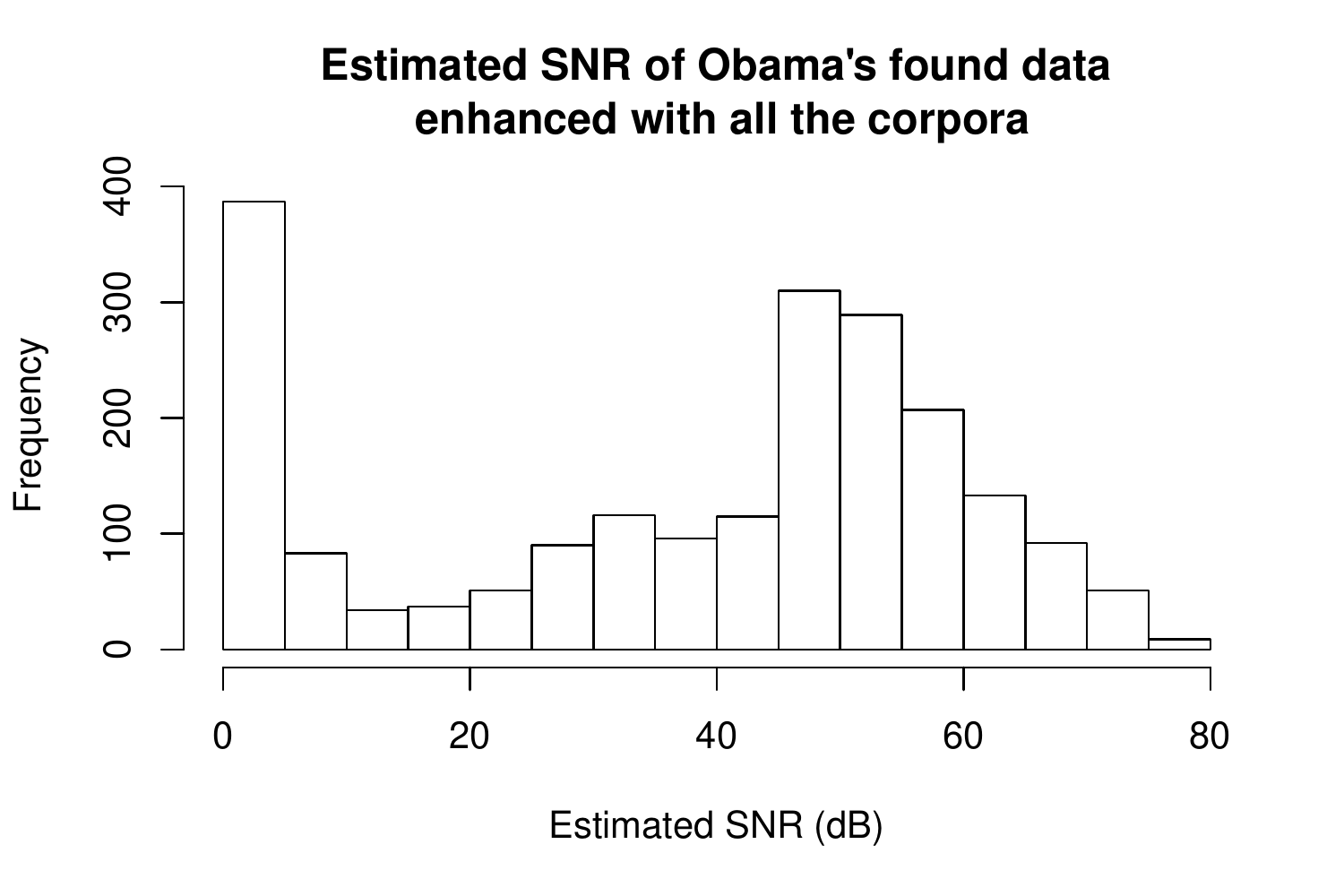}
	\vspace{-5mm}
	\caption{SNR histogram of original Obama voice data after enhancement using all VCTK variants.}
	\label{fig:allhist}
	\vspace{-5mm}
\end{figure}

The SNR estimation results (table~\ref{tab:ObjResults}) show a clear picture: the enhancement process, regardless of which training data were used, improved the average SNR of the original Obama voice data. In particular, training using noisy data (i.e., ``n'', ``nr'', and their mixtures) was considerably more effective than training using the other two possibilities (i.e., ``r'' and ``DR''). This is attributed to the fact that their use reduces the noise levels in the signal, which is what the SNR measure targets. Their use may not actually improve the perceptual quality of the voice signals. Histograms showing the improvement in SNR when ``n'' and when all the variants were used are shown in figures \ref{fig:nhist} and \ref{fig:allhist} respectively.

\subsection{Perceptual evaluation of the speech enhancement}\label{sec:eval1}

As mentioned in the previous section, SNR is most likely not the best measure of enhancement. Since the ultimate objective of our research is to produce high quality synthetic speech, be it through speech synthesis or through voice conversion, it makes sense to evaluate perceptual quality from the viewpoint of human users. We thus carried out a crowdsourced perceptual evaluation with Japanese native listeners. We presented the listeners with a set of 16 screens, each corresponding to one of the eight evaluated conditions (original plus seven enhanced versions) for one of two utterances. The evaluators were given two tasks: 1) rate the perceived quality of each speech sample on a MOS scale and 2) rate the cleanliness of each speech sample (i.e., free from noise, reverberation, and artifacts), also on a MOS scale (with 1 being very noisy speech and 5 being clean speech).

The participants were able to listen to each speech sample as many times as they wanted, but they could not proceed to the next sample until they completed both tasks. They were not allowed to return to previous samples. The samples were selected on the basis of their length, evaluating in the all the utterances (i.e., 530 utterances) between 5.0 and 5.9 seconds long. In total this meant that 265 sets for evaluating all the evaluation utterances, which was done 3 times for a total of 795 sets. The participants were allowed to repeat the evaluation up to 8 times, thereby ensuring that there were at least 100 different listeners. A total of 129 listeners took part in the evaluation (72 male, 57 female).

\subsubsection{Results}
\label{sec:results}

The results of the perceptual evaluation (table~\ref{tab:perceptual}) also show a clear picture. For the original Obama voice data, there was a clear perception of noisiness and related factors (MOS score of 2.42 for cleanliness) even though the perceived quality was reasonably high (3.58). Studio recorded clean speech is normally rated 4.5 on average~\cite{king2017blizzard}, whereas the original Obama voice data was rated 3.5, that is, one point less, most likely due to the poor conditions on which these sources were recorded.

\begin{table}[t!]
\centering
\caption{Results of the perceptual evaluation (MOS score). Non-statistically significant differences are marked with *.}
\vspace{-3mm}
\begin{tabular}{|l|l|l|}\hline
SOURCES & Quality & Cleanliness \\ \hline
Obama source & 3.58* & 2.42 \\
n & 2.73 & 3.35 \\
r & 3.55* & 3.17 \\
nr & 3.11 & 3.42* \\
DR & 3.51 & 3.31 \\
n+DR & 3.26 & 3.02 \\
nr+DR & 3.30 & 3.34 \\
all (n+r+nr+DR) & 3.41 & 3.40*\\ \hline
\end{tabular}
\label{tab:perceptual}
\vspace{-3mm}
\end{table}	

Use of the enhanced versions, improved the cleanliness of the source data, with different degrees of improvement depending on the source data used. Most noteworthy is the cleanliness result for ``noisy-reverberant'' (3.42), which had the largest improvement. This is attributed to the original data being recorded mostly in noisy environments with reverberation, so a speech enhancement system targeting this condition gives the best improvement in that field. The cleanliness result for ``all'' was similarly high, which we attribute to the training being done for all possible situations.

On the other hand, there was a cost to applying these speech enhancements: a consistent degradation in the perceived speech quality. This implies that speech enhancement focused on cleanliness can greatly reduce the naturalness of the speech. This means that the approaches providing the biggest improvements in SNR, such as the ``noisy'' condition with a quality score of 2.73 or the ``noisy-reverberant'' condition with a quality score of 3.11, may not be the best way to produce clean speech for further speech processing.

In short, there seems to be a trade-off between quality degradation and cleanliness improvement, which is not encouraging. But, if we look at the results for the "all" condition, combining all possible data sources, we see that it provided one of the best cleanliness scores (3.40) with one of the smallest quality degradations (0.17 degradation). This strongly suggests that having trained our speech enhancement system in a variety of degradation conditions gave the system enough generalization capability and enough knowledge of human speech to reduce noisiness while maintaining as much as possible voice naturalness.

\section{Generation of the synthetic samples}\label{sec:generation}
\begin{figure}[tb]
  \begin{center}
    \includegraphics[width=0.45\textwidth]{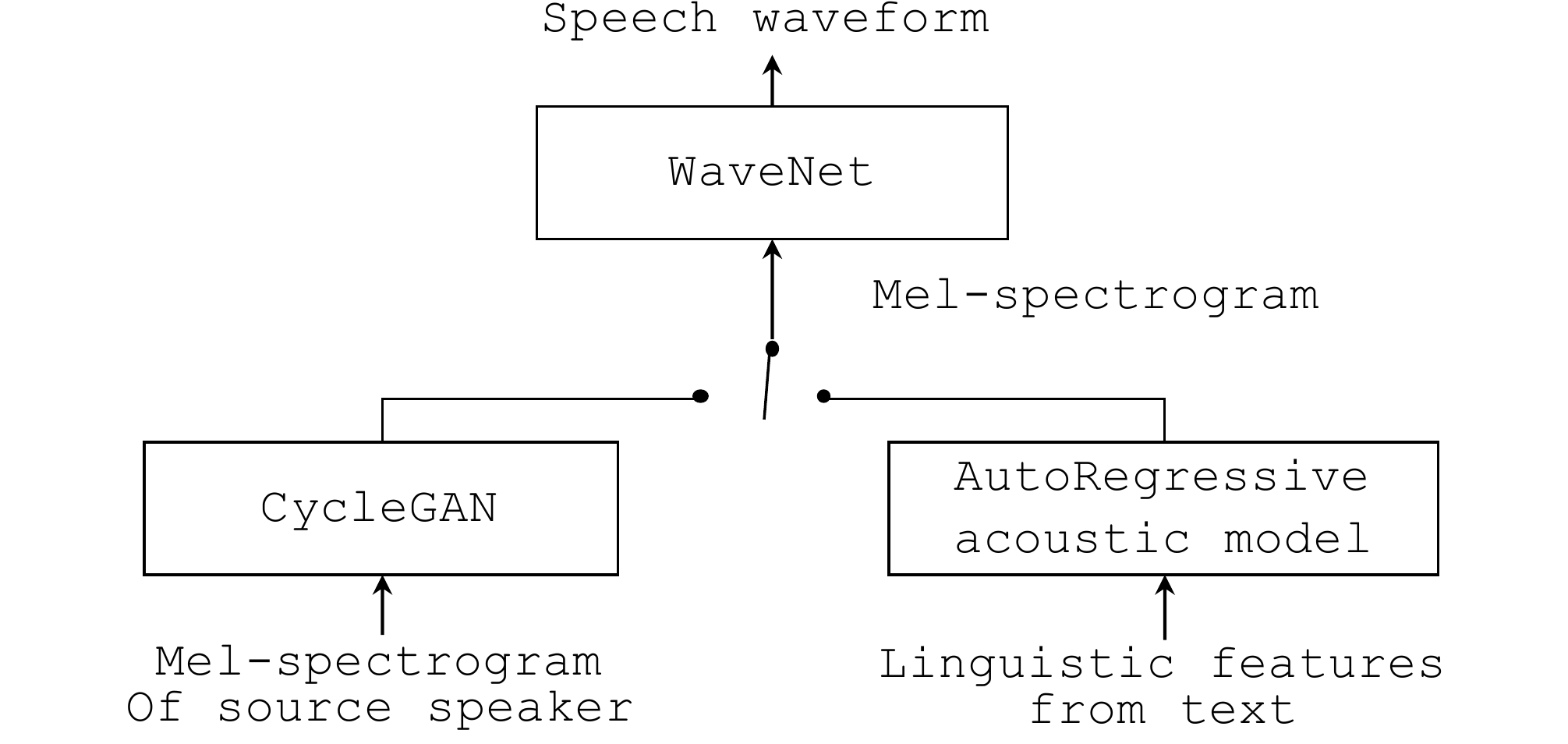}
    \caption{Process for generating spoofed speech.}
    \label{fig:speechgen}
  \end{center}
\end{figure}

We used two approaches to generate spoofed speech waveforms: CycleGAN~\cite{CycleGAN} for VC (section ~\ref{sec:vc}) and an autoregressive (AR) neural network for TTS acoustic modeling (section~\ref{sec:tts}). While both approaches generate mel-spectrograms, CycleGAN converts the mel-spectrogram from a source speaker into a mel-spectrogram that retains the speech contents and overlays the voice characteristics of the target speaker. In contrast, the AR approach converts the linguistic features extracted from text into the mel-spectrogram of the target speaker. Given the generated mel-spectrogram for the target speaker, the WaveNet neural network generates the speech waveforms (section~\ref{sec:wavenet}). The process for generating spoofed speech is illustrated in figure~\ref{fig:speechgen}.

The decision to use the mel-spectrogram as the acoustic feature was based on the expected limitations of traditional features (e.g., F0, Mel-generalized cepstrum, and aperiodicity bands) as the estimation of F0 is problematic in both the original noisy speech signals and the enhanced signals when considering the noisy data we found. We also used an increased number of mel bands compared to other approaches~\cite{46619} (80 vs. 60) with the expectation that it would help the waveform model better cope with corrupted or noisy segments.

\subsection{VC based on CycleGAN}
\label{sec:vc}
\begin{figure}[tb]
  \begin{center}
    \includegraphics[width=0.45\textwidth]{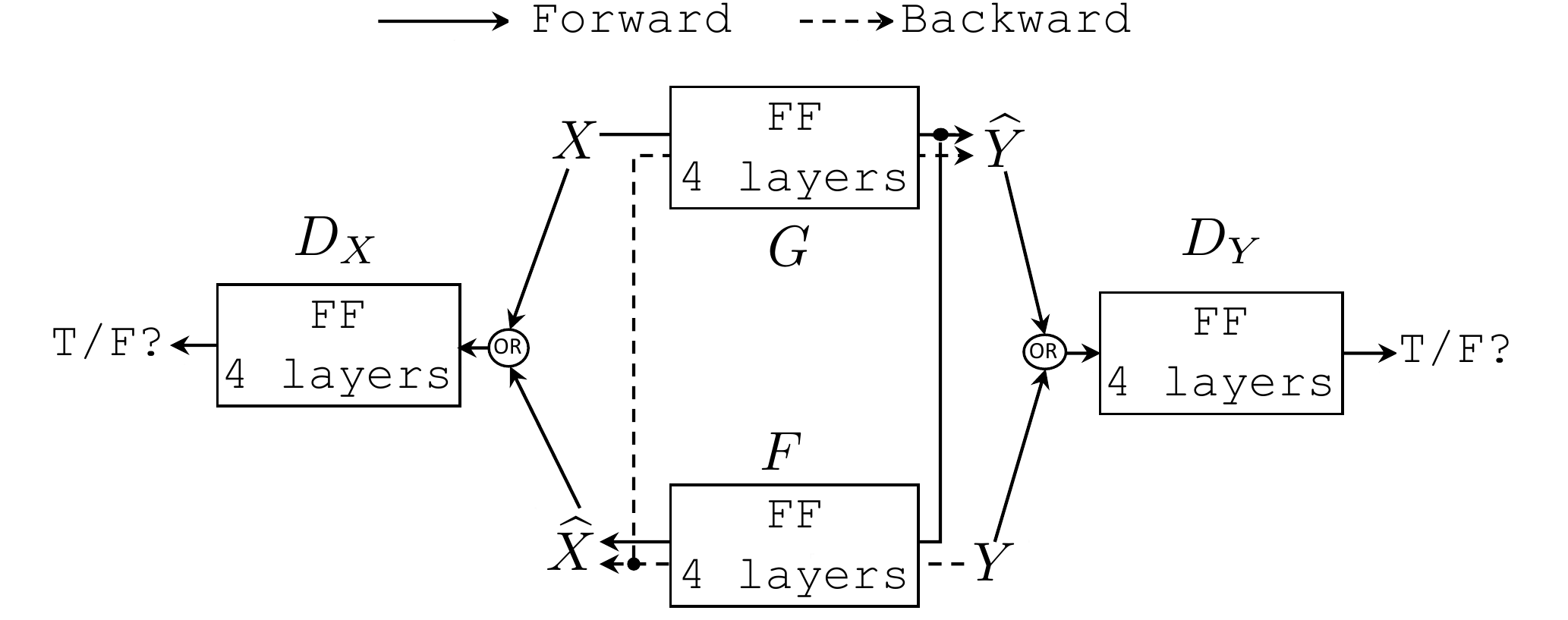}
    \caption{Diagram of CycleGAN. $G$ and $F$ are generators; $D_X$ and $D_Y$ are discriminators. $X$ and $Y$ are real distributions, and $\hat{Y}$ and $\hat{X}$ represent the corresponding generated distributions. `FF' means feed-forward neural network.}
    \label{fig:cyclegan}
  \end{center}
  \vspace{-5mm}
\end{figure}

CycleGAN was originally developed for unpaired image-to-image translation, which consists of two generators ($G$ and $F$) and two discriminators ($D_X$ and $D_Y$), as shown in figure~\ref{fig:cyclegan}. Generator $G$ serves as a mapping function from distribution $X$ to distribution $Y$, and generator $F$ serves as a mapping function from $Y$ to $X$. The discriminators aim to estimate the probability that a sample came from real data $x \in X$ (or $y \in Y$) rather than from the generated sample $\hat{x} = F(y)$ (or $\hat{y} = G(x)$). 

As shown in the same figure, CycleGAN has two translation directions: {\it forward} ($X \rightarrow \hat{Y} \rightarrow \hat{X}$) and {\it backward} ($Y \rightarrow \hat{X} \rightarrow \hat{Y}$). This means that $X$ to $Y$ translation and $Y$ to $X$ translation can be learned simultaneously. Furthermore, an input is carried back to its original form through each translation direction and thereby minimizing consistency loss:
\begin{align}
\mathcal{L}_{cyc} (G, F) &=\EX_{x\sim p_{data}(x)}[\| F(G(x))-x \|_1] \nonumber \\
&+\EX_{y\sim p_{data}(y)}[\|G(F(y))-y \|_1],
\label{eq:cycloss}
\end{align}
where $\EX$ is the expectation and $\| \cdot \|_1$ is the L1 norm. With this structure, it is possible to keep part of the information unchanged when an input is translated by $G$ or $F$. When applying this model to VC, $X$ and $Y$ can be thought of as the feature distribution of the source speaker and that of the target speaker, respectively. By reconstructing the input data, linguistic information may be retained during translation. Additionally, speaker individuality can be changed by adversarial learning using an adversarial loss~\cite{mao2017least}. By integrating consistency loss and adversarial loss, we can learn a mapping function for VC using a non-parallel database~\cite{vc2018cyclegan}.

To train a CycleGAN-based VC system, we used a female speaker (Japanese-English bilingual) as the source speaker. Her speech was recorded in a studio. The target speaker was President Barack Obama\footnote{Audio samples are available at \url{https://fangfm.github.io/crosslingualvc.html}}. Both his original voice data and the enhanced data with SNR $>$ 30 dB (9240 utterances) were used. In accordance with the source speaker's data sets, we implemented three VC systems: one using 611 Japanese utterances, one using 611 English utterances, and one using a mixture of Japanese and English utterances. The generator and discriminator of the CycleGAN were a fully connected neural network with six layers. A 240-dimension vector consisting of a 80-dimension mel-spectrogram and the first and second derivatives were input into the CycleGAN. There were 256, 512, 512, and 256 units in the hidden layers. A sigmoid was used as the activation function. Batch normalization~\cite{ioffe2015batch} was conducted for each hidden layer of the generators. The batch size and learning rate for the generators and discriminators were randomly selected 128 and 4096 frames and 0.001 and 0.0001, respectively.

\subsection{TTS acoustic modeling based on AR neural network}
\label{sec:tts}
An acoustic model for TTS converts the linguistic features extracted from a text into acoustic features such as a mel-spectrogram. Specifically for this work, given a sequence of linguistic feature $\bs{l}_{1:N}=\{\bs{l}_1, \cdots, \bs{l}_N\}$ of $N$ frames, an acoustic model needs to generate a sequence of acoustic features $\bs{a}_{1:N}=\{\bs{a}_1, \cdots, \bs{a}_N\}$ with the same number of frames. Here, $\bs{l}_n$ and $\bs{a}_n$ denote the linguistic features and the mel-spectrogram for the $n$-th frame.

\begin{figure}[!t]
  \begin{center}
    \includegraphics[width=0.5\textwidth]{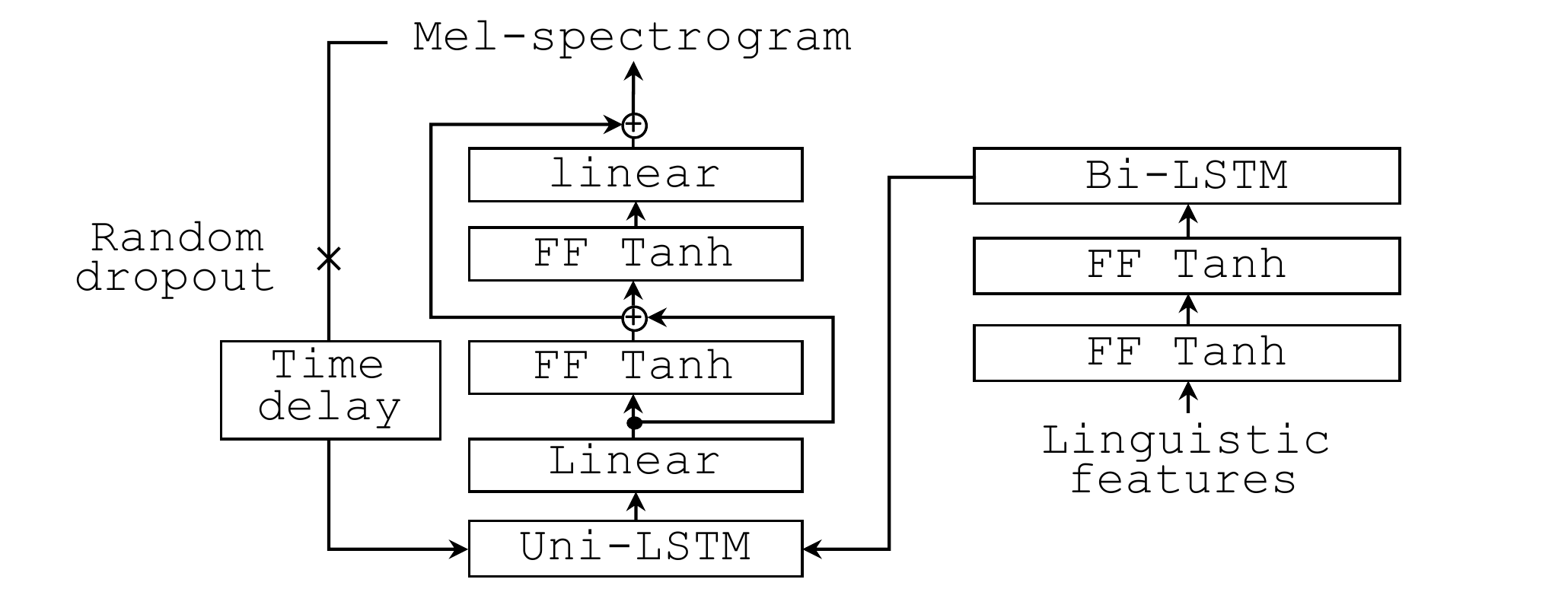}
    \vspace{-5mm}
    \caption{AR neural network for TTS acoustic modeling. FF Tanh and Linear denote feedforward layers with Tanh and identity activation function, respectively. Bi-LSTM and Uni-LSTM denote bi-directional and uni-directional LSTM layers. The ``Time delay'' block keeps the Mel-spectrogram for frame $n$ and sends it to the Uni-LSTM at frame $n+1$.}
    \label{fig:ARmodel}
  \end{center}
  \vspace{-5mm}
\end{figure}
The AR neural network illustrated in figure~\ref{fig:ARmodel} was used to convert $\bs{l}_{1:N}$ into $\bs{a}_{1:N}$. This network has two feedforward layers and a bi-directional long-short-term-memory (LSTM) unit recurrent layer near the input side. Following these three layers, it uses another uni-directional LSTM layer. Different from the first LSTM layer, this one not only takes the output of the previous layer but also the previous output of the whole network as input. For example, it takes $\bs{a}_{n-1}$ as input when it generates $\bs{a}_{n}$ for the $n$-th frame. This type of data feedback is widely used in neural text generation and machine translation~\cite{sutskever2014sequence,bengio2015scheduled}, and a network with this type of feedback loop is referred to as an autoregressive model. Note that, while the natural $\bs{a}_{n-1}$ from the training set is fed back during training, the generated $\widehat{\bs{a}}_{n-1}$ is fed back during generation. Also note that a simple trick is used here that improves network performance: it is to randomly drop out the feedback data in both the training and generation stages~\cite{wangDARF0}. 

The linguistic features used for both training and synthesis were extracted using Flite~\cite{HTSWorkingGroup2014}. The dimension of these feature vectors was 389. The alignment information was obtained using forced alignment with a hidden-semi Markov model trained using HTS~\cite{ref:HTS-NITECH-2005-IEICE} on the mel-spectrograms. In addition to the linguistic features, a numeric variable characterizing the enhancement condition was used as input. Together with the input features, the network was trained using the mel-spectrograms obtained with the enhanced speech method explained in section~\ref{sec:enhcobamac}. The dropout rate was set to 25\%.

\subsection{WaveNet vococder}
\label{sec:wavenet}	

Building a state-of-the-art data-driven vocoder such as WaveNet represents a big challenge when trying to use the types of data we found: it is not easy to gather sufficient data that are good enough for the process. This is where the advantage of having used another data-driven speech enhancement system comes into play. As hinted at in the introduction to section~\ref{sec:enhcobamac}, we can take advantage of our GAN-based speech enhancement system to generate multiple enhanced versions of the noisy speech data, effectively multiplying the amount of training data available for training our system.

\begin{figure}[!t]
  \begin{center}
    \includegraphics[width=0.45\textwidth]{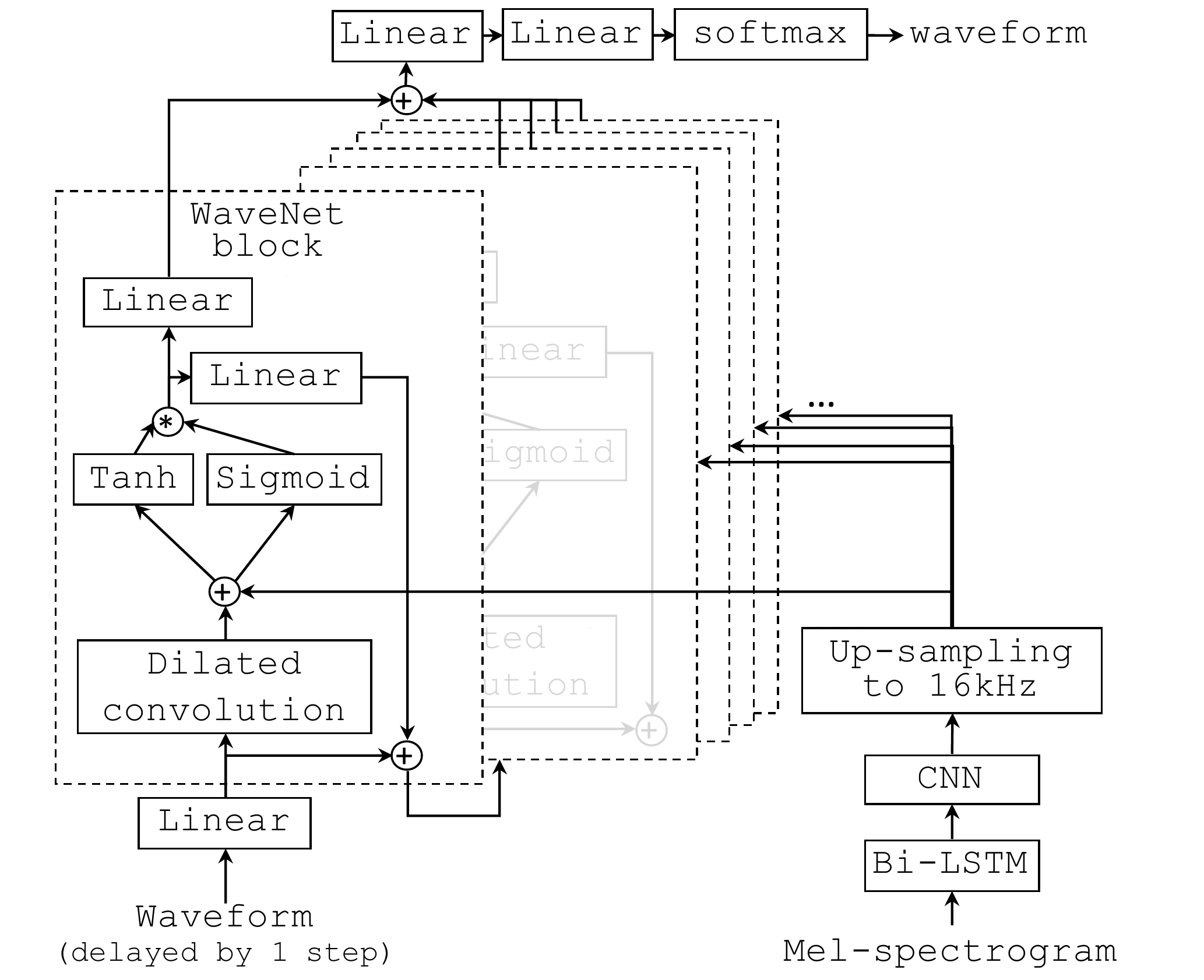}
    \vspace{-2mm}
    \caption{Structure of WaveNet vocoder. Each WaveNet block is surrounded by dotted line. Details of first WaveNet block are shown.}
    \label{fig:WaveNet}
  \end{center}
  \vspace{-5mm}
\end{figure}

For this research, we trained our WaveNet vocoder on the enhanced version of the original Obama voice data. The network structure is illustrated in figure~\ref{fig:WaveNet}. The vocoder works at a sampling rate of 16 kHz. The $\mu$-law compressed waveform is quantized into ten bits per sample. Similar to that in a previous study~\cite{Tamamori2017}, the network consists of a linear projection input layer, 40 WaveNet blocks for dilated convolution, and a post-processing block. The $k$-th dilation block has a dilation size of $2^{\mathrm{mod}(k,10)}$, where $\mathrm{mod}(\cdot)$ is modulo operation. In addition, a bi-directional LSTM and a 1D convolution layer are used to process the input acoustic features. The acoustic features fed to every WaveNet block contain the 80-dimensional mel-spectrogram plus an additional component specifying which of the different speech-enhancing models produced that speech waveform\footnote{The tools for the TTS and WaveNet implementation are based on a modified CURRENNT toolkit~\cite{weninger2015introducing} and can be found online http://tonywangx.github.io}. 

\section{Perceptual evaluation of the generated speech}

To evaluate the generation capabilities of the proposed system, we carried out a second crowd-sourced perceptual evaluation with Japanese native listeners. The evaluation was conducted in the same way as the one described in section~\ref{sec:eval1}. Only the tasks were different: 1) rate the quality of the voice in each speech sample, ignoring the effects of background noise and 2) rate the similarity of the voice to that of the spoofed speaker (i.e., Obama), ignoring speech quality and cleanliness. To compare with the target speaker, we presented the participants two additional samples (one generated by the evaluated system and one with natural speech), both containing the same linguistic content but different than the one used for rating speech quality. The participants were asked to rate on a MOS scale the similarity of the two speakers.

\subsection{Systems evaluated}
We evaluated six systems plus natural speech and copy synthesis of the mel-spectrograms of the natural speech with the trained WaveNet vocoder:
\begin{itemize}
\item Voice conversion system 1 (VC1): Japanese-to-English bilingual VC
\item Voice conversion system 2 (VC2): English-to-English monolingual VC
\item Voice conversion system 3 (VC3): Japanese and English mixed utterances multilingual VC
\item Text to speech system 1 (TT1): text to speech from estimated mel-spectrogram for noisy condition
\item Text to speech system 2 (TT2): text to speech from estimated mel-spectrogram for enhanced-all condition
\item Text to speech system 3 (TT3): Speech generated with TT2 with a small added amount of reverberation
\end{itemize}
With the three different voice conversion systems, we aimed to evaluate the effect of noise and to investigate whether CycleGAN can be used for cross-lingual VC systems. 
With the three TTS systems, we aimed to analyze the effect of generating spectrograms on the basis of different conditions for speech enhancement and to determine the importance of mimicking the environmental conditions of the reference natural speech when considering human perception.

\subsection{Results}
\label{sec:results2}

\begin{table}[t!]
\centering
\caption{Results of second perceptual evaluation (MOS score). Non-statistically significant differences are marked with superscripts.}
\vspace{-3mm}
\begin{tabular}{|l|l|l|}\hline
Sources & Quality & Similarity \\ \hline
Natural & 4.40 & 4.70 \\
Copy-synthesis & 2.45$^1$ & 2.99 \\
VC1 & 2.66$^2$ & 1.56$^3$ \\
VC2 & 2.67$^2$ & 1.55$^3$ \\
VC3 & 2.83 & 1.56$^3$ \\
TT1 & 2.49$^1$ & 1.43$^4$ \\
TT2 & 2.51$^1$ & 1.40$^4$ \\
TT3 & 2.63$^2$ & 1.45$^4$ \\ \hline
\end{tabular}
\label{tab:perceptual2}
\vspace{-3mm}
\end{table}	

The results for the second perceptual evaluation are summarized in table~\ref{tab:perceptual2}. There was a total of 103 unique participants (52 male and 51 female).
The results for natural speech indicate that the participants were able to identify the actual Obama voice regardless of the environmental conditions (MOS score of 4.70 for similarity). They also indicate that they were able to distinguish the naturalness and frequency of the speech regardless of the background noise and/or reverberation (MOS score of 4.40 for quality).

The results for copy synthesis using the trained WaveNet vocoder were quite different (MOS score of 2.45 for quality). The WaveNet system and scripts had previously been successful at generating speech when clean speech was used for training~\cite{xin2018icassp}, suggesting that the difference was due to the nature of the data used for training. One possibility is that using mixture density networks for generating output waveforms is problematic for modeling the variance of noisy data.

Looking at the TTS system results, we see that the quality of the generated speech did not change significantly with the generation condition (MOS score of ~2.5) and was similar to that for copy synthesis. Adding a small amount of reverberation improved the perceived speech quality so that it was even higher than with copy synthesis (2.63 versus 2.45). This means that reverberation can mask part of the noise generated by the WaveNet vocoder. The significant drop in the similarity score means that we cannot say whether the evaluators were capable of identifying Obama's voice.

Looking at the VC system results, we see a similar pattern. In terms of quality, the VC systems were slightly but significantly better than both TTS and copy synthesis. This is probably because the VC systems were trained on selected data (i.e., only data with estimated SNR $>$ 30 dB), and clean data was used for the source speaker. In terms of similarity, they were slightly but significantly better than TTS but far worse than copy synthesis. 

Comparing VC1 (Japanese-to-English bilingual VC) with VC2 (English-to-English mono-lingual VC), we see that they achieved similar MOS scores for both speech quality and speaker similarity. This suggests that CycleGAN can be used to train a bilingual VC model. When Japanese and English utterances were mixed (VC3), the speech quality was slightly higher for the other VC systems. This is probably because twice the amount of source speaker training data was used.

\section{Evaluation based on anti-spoofing countermeasures} 

We have also evaluated the built TTS and VC systems based on anti-spoofing countermeasures. The countermeasure used is a common Gaussian mixture models (GMMs) back-end classifier with constant Q cepstral coefficient (CQCC) features \cite{Todisco+2016}. Training of the countermeasure is speaker-independent and we have used two alternative training sets to train the two GMMs (one for natural or bona fide speech; another one for synthetic or converted voices). The first one contains the training portion of the ASVspoof2015 data consisting of 5 (now outdated) spoofing attacks, and the second one consists of the converted audio samples submitted by the 2016 Voice Conversion Challenge (VCC2016) participants \cite{Toda+2016}. The latter contains 18 diverse stronger VC attacks\footnote{http://dx.doi.org/10.7488/ds/1575}. 

Table 7 shows the evaluation results for the CQCC-GMM countermeasure when scored on the found data of Obama (that is, the same data used for the listening tests in a previous section). The results are presented in terms of equal error rate (EER, \%) of the spoofing countermeasure. The higher the EER, the more confused the countermeasure will be in telling apart our generated voices from natural human speech.

As we can see from a table~\ref{tab:eer}, although the VC and TTS systems in this paper are more advanced methods than ones included in the current ASVspoof 2015, the countermeasure models can still detect both the proposed VC and TTS samples using the found data easily. This is because not only the VC and TTS process but also the additional speech enhancement process  caused noticeable artifacts.  

\begin{table}[t!]
\centering
\caption{Evaluation results based on anti-spoofing countermeasures (EER in percentages). 32-mix CQCC-GMMs were trained on ASVspoof2015 or VCC2016 sets.}
\vspace{-3mm}
\begin{tabular}{|l|l|l|}\hline
Sources & ASVspoof2015 & VCC2016  \\ \hline
Copy-synthesis & 4.63 & 8.46 \\
VC1 & 2.32 & 1.08 \\
VC2 & 2.16 & 0.00 \\
VC3 & 2.25 & 1.01 \\
TT1 & 1.60 & 0.00 \\
TT2 & 2.01 & 0.00 \\
TT3 & 0.79 & 0.00 \\ \hline
\end{tabular}
\label{tab:eer}
\vspace{-3mm}
\end{table}

\section{Conclusions and future work}
\label{sec:conc}

We have introduced a number of publicly available and known datasets that proved to be extremely useful for training speech enhancement models. Application of these models to a corpus of low-quality considerably degraded data found in publicly available sources significantly improved the SNR of the data.

A perceptual evaluation revealed that the models can also significantly improve the perceptual cleanliness of the source speech without significantly degrading the naturalness of the voice as is common when speech enhancement techniques are applied. Speech enhancement was most effective when the system was trained using the largest amount of data available as doing so covered a wide variety of environmental and recording conditions, thereby improving the generalization capabilities of the system.

A second perceptual evaluation revealed that, while generating synthetic speech from noisy publicly available data is starting to become possible, there are still obvious perceptual problems in both text-to-speech and voice conversion systems that must be solved to achieve the naturalness of systems trained using very high quality data. Therefore, we cannot recommend yet that next-generation ASVspoof data be generated using publicly available data even if adding this new paradigm of speech generation systems is a must.

\noindent\textbf{Acknowledgements:}
This work was partially supported by MEXT KAKENHI Grant Numbers (15H01686, 16H06302, 17H04687).

\bibliographystyle{IEEEtran}
\bibliography{thesisbib}\label{sec:refs}

\end{document}